\documentclass[journal]{IEEEtran}
\usepackage{xr-hyper} 
\usepackage{hyperref}

\usepackage[acronym,nonumberlist,nogroupskip,style=super]{glossaries}
\usepackage{hyperref}
\newacronym{iot}{IoT}{Internet of Things} %{label}

\usepackage{cite}
\usepackage{amsmath,amssymb,amsfonts}
\usepackage{graphicx}
\usepackage{textcomp}
\usepackage{bmpsize}
\usepackage[dvipsnames]{xcolor}
\usepackage{lipsum}

\usepackage[export]{adjustbox}
\usepackage{pifont}
\usepackage{multirow}
\usepackage{tabularx}
\usepackage{algorithmicx}
\usepackage[noend]{algpseudocode}
\usepackage[T1]{fontenc}
\usepackage{algorithm} 
\usepackage{subfig}	
\usepackage{soul}
\usepackage{booktabs}

\usepackage{multirow}
\usepackage{rotating}
\usepackage{booktabs}

\usepackage{siunitx,etoolbox}

\usepackage{color, colortbl}

\usepackage{comment}
\usepackage{fontawesome5}
\usepackage{orcidlink}
\usepackage[switch, modulo]{lineno}

\begin{document}

\title{Individual Packet Features are a Risk to Model Generalisation in ML-Based Intrusion Detection}

\author{Kahraman~Kostas~\orcidlink{0000-0002-4696-1857},
	Mike~Just~\orcidlink{0000-0002-9669-5067},
	and~Michael~A.~Lones~\orcidlink{0000-0002-2745-9896}% <-this % stops a space
\thanks{K.~Kostas, M.~Just,	and~M.~A.~Lones  are with the Department of Computer Science, Heriot-Watt University, Edinburgh EH14 4AS, UK,  e-mail: kk97, m.just, m.lones@hw.ac.uk}
\thanks{K. Kostas supported by Republic of Turkey Ministry of National Education.}}% 
\maketitle

\begin{abstract}

Machine learning is increasingly used for intrusion detection in IoT networks. This paper explores the effectiveness of using individual packet features (IPF), which are attributes extracted from a single network packet, such as timing, size, and source-destination information. Through literature review and experiments, we identify the limitations of IPF, showing they can produce misleadingly high detection rates. Our findings emphasize the need for approaches that consider packet interactions for robust intrusion detection. Additionally, we demonstrate that models based on IPF often fail to generalize across datasets, compromising their reliability in diverse IoT environments.

\end{abstract}


\section{Introduction}

The continually growing number of connected devices, coupled with the heterogeneity of hardware and software designs across manufacturers, presents a significant challenge for securing the Internet of Things (IoT) ecosystem~\cite{harbi2021recent}. The inherent variety in these devices can lead to complex security vulnerabilities, potentially compromising network security. For example, weakly secured IoT devices can be exploited by malicious actors, turning them into ``botnet zombies'' used in large-scale cyberattacks against critical infrastructure~\cite{kambourakis2017mirai}.

The multifaceted nature of IoT security challenges has attracted considerable scholarly attention. Notably, intrusion detection employing machine learning (ML) techniques has emerged as a particularly active area. Although specific ML methods have evolved over time, the fundamental reliance on data remains constant, underscoring the indispensability of robust datasets in data-driven approaches.

Current intrusion detection studies with ML primarily rely on three feature types~\cite{kostas2023iotgem}: flow-based features (analyzing network traffic statistics)~\cite{sharafaldin2018toward,zhao2013botnet}, window-based features (focusing on packet variations within a timeframe)~\cite{mirsky2018kitsune,kostas2023iotgem}, and individual network packet features~\cite{Anthi2019a,DeCarvalhoBertoli2021}. 

The third approach is the simplest --- it uses features extracted from individual packets, and does not take into account interactions between packets --- yet despite this, previous studies have reported high detection accuracy using this approach.

In this study, we critically examine the efficacy of IPF in ML-based intrusion detection systems, revealing the approach to have significant limitations, particularly in terms of the ability of models to generalise. We demonstrate, through extensive experiments on various public datasets, that reliance on IPF can lead to vulnerabilities and reduced robustness in intrusion detection. Our findings underscore the importance of adopting more informative feature sets, such as flow-based and window-based features, to enhance the security of IoT ecosystems. Notably, this study is the first to provide empirical evidence of the risks associated with IPF.

\section{Why IPF Does Not Tell the Whole Story} \label{why}

Detecting an attack using IPF alone is challenging. In this section, we'll elaborate on the reasons behind this difficulty.  In the subsequent sections, we support this notion by presenting examples from existing literature and our own experimentals. For reproducibility, our scripts are publicly available\footnote{Source code available at:\href{https://github.com/kahramankostas/IPF}{github.com/kahramankostas/IPF}}.

\begin{figure}[htbp]	\centering	\caption{Legitimate and malicious use of 3-way handshake}	\subfloat[\label{fig:syn1}\centering The legitimage usage.]{{\includegraphics[width=50mm]{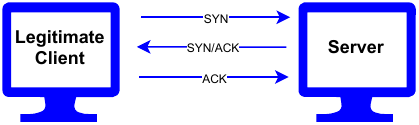}}}\\	\subfloat[\label{fig:syn2}\centering SYN Flood attack.]{{\includegraphics[width=75mm]{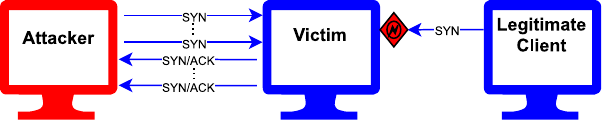}}}\\	\label{fig:synfig}	\end{figure}

This challenge can be explained with a simple example: SYN Flood~\cite{gulihar2020cooperative}, which is a Denial-of-Service  (DoS) attack based on the exploit of 3-way handshake in the TCP protocol. During the 3-way handshake process, the client that wants to establish a connection sends a SYN packet to the server. The server receiving this packet sends a SYN-ACK packet to the client. Finally, the client responds with the ACK packet, and the TCP connection is established.  (see Figure~\ref{fig:synfig}).
During an attack, the server is sent a larger number of SYN packets for which a SYN-ACK reply is sent, but the ACK reply is never received. Server resources are impacted, and subsequent legitimate connection requests could be dropped. The distinguishing individual packet features in this case are typically those such as IP or MAC addresses (or other network-specific features) which are identifying within one dataset, but which do not generalise to other datasets or network environments.

This phenomenon extends beyond SYN Flood attacks; vari-
ous attack scenarios causing time, size, or protocol anomalies,
such as DoS, distributed DoS (DDos) and man-in-the-middle (MitM) attacks, display similar characteristics.
To efficiently detect and mitigate an attack, it is essential to
adopt an approach that considers the context and
interaction of packets, rather than solely relying on individual
packet characteristics.

There are some exceptions, such as for single-packet attacks that exploit malformed packets. These attacks typically involve sending malformed packets that the device cannot handle, leading to malfunctions or crashes in the receiving device~\cite{singlepackets}. However, addressing these attacks usually involves packet filtering or firewalls, not ML behavioural analysis, since they can be effectively managed using signature/rule-based approaches.

\section{IPF Usage in the Literature} \label{lr}

The use of individual packets in attack detection is common. Our literature review, conducted in ~\cite{kostas2023iotgem}, reveals that 10 out of 68 studies from 2019 to 2023 incorporate IPFs. Table~\ref{tab:ind-results} summarises key information from these studies, such as the dataset, data types, and metrics, and Table~\ref{tab:attacks} lists the attacks detected.

This reveals numerous studies with reported high success rates in attack detection using IPF. Given that we have posited that IPFs are an ineffective basis for attack detection, it is important to understand this discrepancy.

\begin{table}[htbp]
	\centering
	\caption[Studies in the literature using IPF.]{Studies in the literature using IPF, their datasets and results. The dataset specified as Private* is the same in both studies.}
	\resizebox{0.5\textwidth}{!}{\begin{tabular}{@{}llrrrr@{}}
		\toprule
		Study & Dataset & \multicolumn{1}{l}{Accuracy} & \multicolumn{1}{l}{Recall} & \multicolumn{1}{l}{Precision} & \multicolumn{1}{l}{F1 Score} \\
		\midrule
		
		\cite{ghourabi2022security} & \href{https://ieee-dataport.org/documents/edge-iiotset-new-comprehensive-realistic-cyber-security-dataset-iot-and-iiot-applications}{Edge-IIoTset}(CSV)  & 100.00   & 100.00   & 100.00   & 100.00 \\
		\cite{ferrag2022edge} & \href{https://ieee-dataport.org/documents/edge-iiotset-new-comprehensive-realistic-cyber-security-dataset-iot-and-iiot-applications}{Edge-IIoTset}(CSV)  & 100.00   & 89.00    & 95.00    & 87.00 \\
		\cite{saran2023comparative} & \href{https://ieee-dataport.org/open-access/mqtt-iot-ids2020-mqtt-internet-things-attack-detection-dataset}{MQTT-IoT-IDS2020}(CSV) & 99.98 & 99.98 & 99.98 & 99.98 \\
		\cite{friha2022felids} & \href{https://www.kaggle.com/datasets/cnrieiit/mqttset}{MQTTset}(CSV) & 91.00    & 91.00    & 77.00    & 80.00 \\
		\cite{DeCarvalhoBertoli2021} & \href{https://github.com/c2dc/AB-TRAP/tree/main/}{AB-TRAP} (CSV) &       &       &       & 100.00 \\
		\cite{chen2022intrusion} & \href{https://icsdweb.aegean.gr/awid/}{AWID2} (RAW) & 99.96 & 99.99 & 99.95 & 99.97 \\
		\multirow{1}[0]{*}{\cite{han2023network}} & \href{https://www.unb.ca/cic/datasets/ids.html}{ISCXIDS2012} (RAW) & 99.42 & 99.41 & 99.34 & 99.37 \\
		& \href{https://www.unb.ca/cic/datasets/ids-2017.html}{CICIDS2017} (RAW) & 97.87 & 98.16 & 97.59 & 97.83 \\
		
		\cite{mondal2022comparative} & Private & 100.00   & 100.00   & 100.00   & 100.00 \\
		
		\cite{Anthi2019a} & Private* &       & 98.00    & 99.00    & 99.00 \\
		\cite{Anthi2021} & Private* &       & 100.00   & 100.00   & 100.00 \\
		\bottomrule
		
	\end{tabular}}%
	
	\label{tab:ind-results}%
\end{table}%

% Table generated by Excel2LaTeX from sheet 'Sheet3'
\begin{table}[htbp]
  \centering
  \caption{List of studies and the attacks they include}
    \begin{tabular}{@{}ll@{}}
    \toprule
    Study & Attacks \\
    \midrule
    \cite{ghourabi2022security}  & DoS/DDoS, Scanning, MitM, Injection, Malware \\
    \cite{ferrag2022edge}  & DoS/DDoS, Scanning, MitM, Injection, Malware \\
    \cite{saran2023comparative}  & Brute-Force,  Scanning,  \\
    \cite{friha2022felids}  & DoS/DDoS, MitM, Injection, Packet Manipulation  \\
    \cite{DeCarvalhoBertoli2021}  & Scanning \\
    \cite{chen2022intrusion}  & DoS, Injection,  Authentication ,  Wireless Attack  \\
    \cite{han2023network} & DoS/DDoS, Brute-Force, Infiltration, Scanning, Botnet, Web \\
    \cite{mondal2022comparative}  & Wrong setup, DDoS, Probing, Scanning, MitM \\
    \cite{Anthi2019a}  & DoS, MitM, Scanning, IoT-toolkit \\
    \cite{Anthi2021}  & DoS \\
    \bottomrule
    \end{tabular}%
  \label{tab:attacks}%
\end{table}%

Most studies in Table~\ref{tab:ind-results} use open access datasets, and those used in~\cite{DeCarvalhoBertoli2021,ghourabi2022security,	ferrag2022edge,	saran2023comparative,friha2022felids} have a similar nature in that all contain raw data as well as individual packet specifications in pre-extracted CSV format. All of the studies used these pre-extracted versions. Another feature of these datasets is that they contain a single file for each attack, with no information about sessions. Therefore, in all these studies, it seems likely that the training and test data were created from a single file.

\begin{figure}[htbp]
	\centering
	\caption[An 80/20 split for a hypothetical dataset containing 40 samples.]{An 80/20 split for a hypothetical dataset containing 40 samples (4 session, 40 network packets), showcasing the division into 80\% training data and 20\% testing data for model development and evaluation, respectively. This example illustrates how information leakage takes place. In the scenario involving the source IP attribute, information leakage occurs because this address uniquely identifies both the attacker and the benign source, remaining constant across both the training and test sets. If we examine other features such as source port numbers and ID numbers, they vary across different sessions but exhibit correlations within each session (port numbers remain constant, ID numbers increment sequentially). Consequently, if packets from the same session are found in both the training and test sets, it can result in information leakage. Benign: \textcolor[HTML]{d5e7d4}{\faCircle}
		\textcolor[HTML]{cceb8b}{\faCircle} Malicious: \textcolor[HTML]{f7cdcc}{\faCircle}
		\textcolor[HTML]{ff9999}{\faCircle}}
	\centering{\includegraphics[width=90mm]{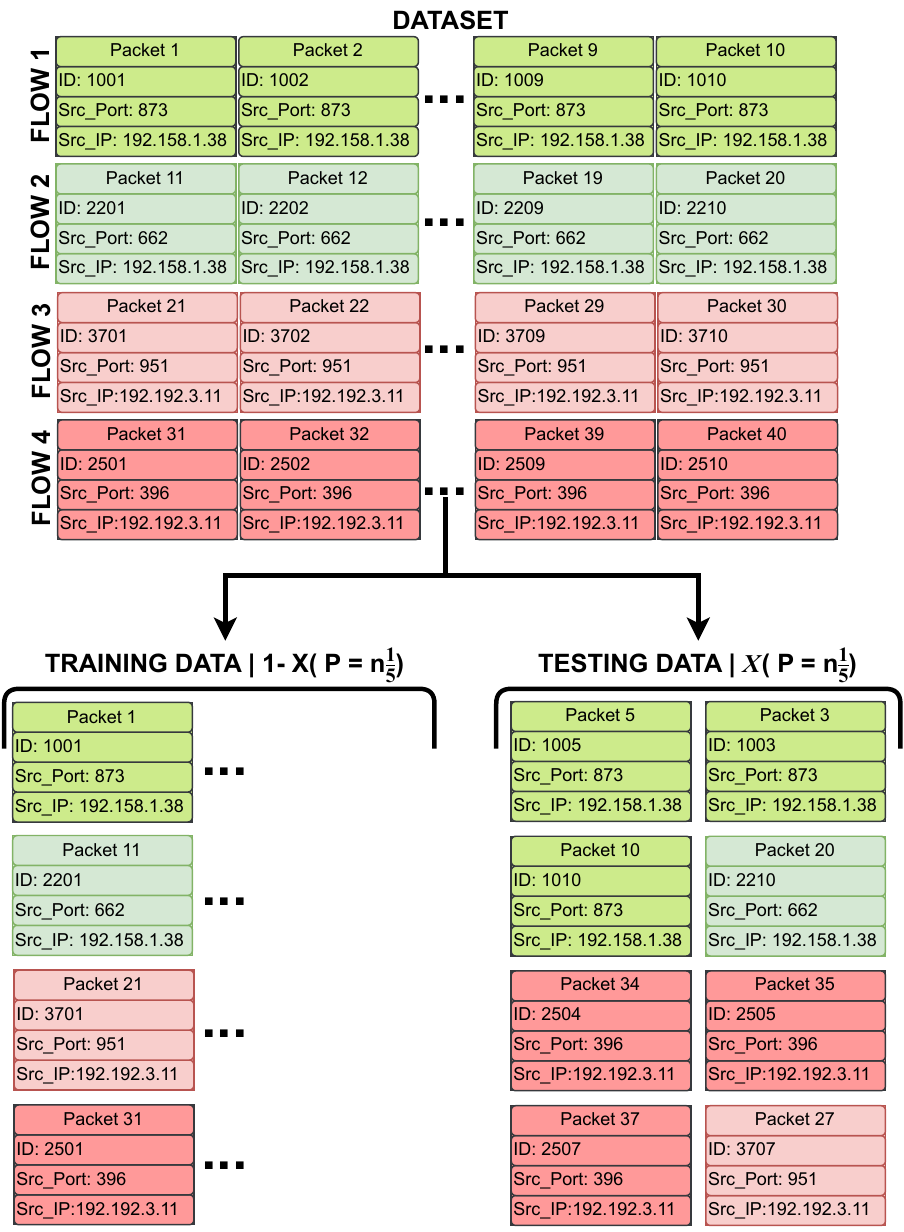}}%
	\label{fig:packetflow}
\end{figure}

This promotes identifying features, such as IP addresses, across training and test sets, causing information leakage. Such leakage can inflate performance metrics by granting the model access to information during testing that it would not have during deployment. Consequently, these features serve as hidden variables, offering a shortcut to class identification. The model may then prioritize them over learning true underlying patterns, as illustrated in Figure~\ref{fig:packetflow}. For example, in~\cite{ghourabi2022security,saran2023comparative,friha2022felids}, although the reporting of features used in these studies is unclear, there is no evidence that these identifying features have been removed or censored. In this respect, it is quite possible that identifying features were used in their models.

Nevertheless, the inclusion of source and destination-based identifiers is generally considered a significant flaw, and most studies tend to remove such features from their data.  For instance, in~\cite{ferrag2022edge}, IP addresses were removed. However, this study suffers from a different type of information leak, with the dataset containing features such as ports, IDs, synchronisation, and acknowledgement numbers. Despite these features being extracted from individual packets, they provide non-generalizable insights into the interrelation of packets. For instance, when a flow is established between two ports, all individual packets within that flow bear the corresponding port numbers (see Figure~\ref{fig:packetflow} as an example, focusing on port and ID numbers). In a train/test split that does not consider the flows, even with more robust evaluation methods like cross-validation (CV), it is highly likely that packets representing the same flow will occur in both the training data and the test data. Moreover, the fact that a single file with no distinct sessions is used to represent each attack makes it very challenging to split in a manner that prevents information leakage. So, in this study, it can be seen that the features that stand out for many attacks are sequence and acknowledgement numbers, which are session-based identifiers, and it is quite possible that they are subject to information leakage.

In none of these studies~\cite{DeCarvalhoBertoli2021,ghourabi2022security,	ferrag2022edge,	saran2023comparative,friha2022felids} did we see any indication that flows were taken into account in the training and test split, or that these session-based identifiers were removed  (Although session-based features are mostly removed in Bertoli et al.'s work~\cite{DeCarvalhoBertoli2021}, the low complexity in the structure of the dataset still leads to generalisation problems. this will be examined in Section~\ref{low}) .  Therefore information leakage between the training and test data is distinctly possible. While the results published in these studies may appear promising, models relying on identifying features due to information leakage will not generalise to other datasets, and consequently the results are likely to be misleading.

Two studies \cite{chen2022intrusion,han2023network} used raw pcap files rather than pre-extracted features. 
However, the use of raw data can be quite dangerous because these files also contain identifiers such as IP and MAC addresses. In one of these studies \cite{chen2022intrusion}, there is no indication that these identifying features were deleted/discarded, so it is quite likely that they were used.
In the other study\cite{han2023network}, 
identifying features such as IP addresses and frame number were removed. Despite this precaution, we suspect that some identifying information might have inadvertently leaked. For example, in their study, the internet checksum has been used, which is an identifying feature as it stores a hashed version of the source and destination IP addresses. In addition, both studies had no mechanism to prevent sessions from being split across train and test sets, providing a potential route for information leakage.

The other three studies~\cite{mondal2022comparative,Anthi2019a,Anthi2021} listed in Table~\ref{tab:ind-results} did not use publicly available data. Many of the features used in \cite{mondal2022comparative} are identifying features (such as frame number/time, eth source/destination, source/destination IP). 

In two studies~\cite{Anthi2019a,Anthi2021}, identifying features such as IP, MAC, and frame number were removed, but session-based segregation was not applied, likely causing session characteristics to leak between training and test sets. Notably, \cite{Anthi2019a} claims the testing phase was isolated from training data, although the process for achieving this is not clear.

\section{Experimental Case Studies}

In the previous section, we highlighted the  role that identifying features may play in giving an incorrect impression that IPF-based approaches are effective. Next, we strengthen this claim using experimental case studies. We also consider another potentially misleading factor, low data complexity.

\subsection{Identifying Features}
\begin{figure*}[h]
	\centering

	\caption[Comparison of CV and isolated data.]{Comparison of CV and isolated data performance of features on some attacks in the IoT-NID dataset with DT.}
	\subfloat[\label{fig:attack1}\centering IoT-NID dataset ARP Spoofing attack.]{{\includegraphics[width=80mm]{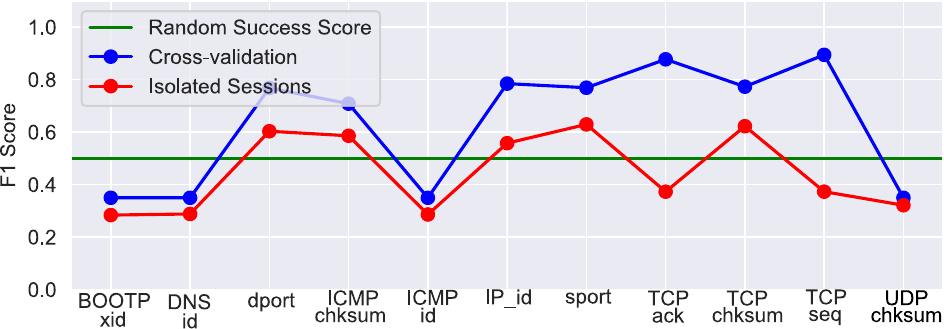}}}
	\subfloat[\label{fig:attack2}\centering IoT-NID dataset Telnet Bruteforce attack.]{{\includegraphics[width=80mm]{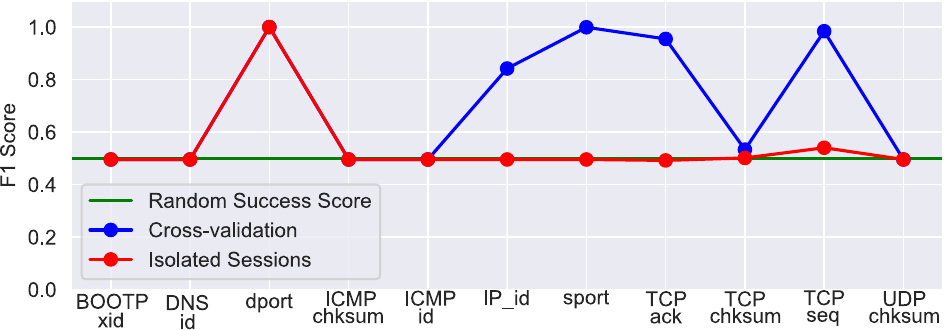}}}\\
	\subfloat[\label{fig:attack3}\centering IoT-NID dataset HTTP flooding attack.]{{\includegraphics[width=80mm]{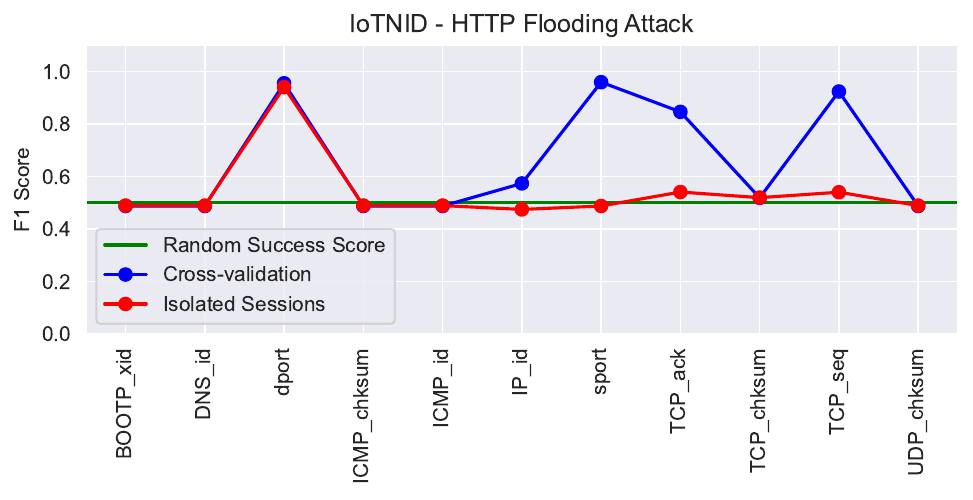}}}
	\subfloat[\label{fig:attack4}\centering IoT-NID dataset UDP Flooding attack.]{{\includegraphics[width=80mm]{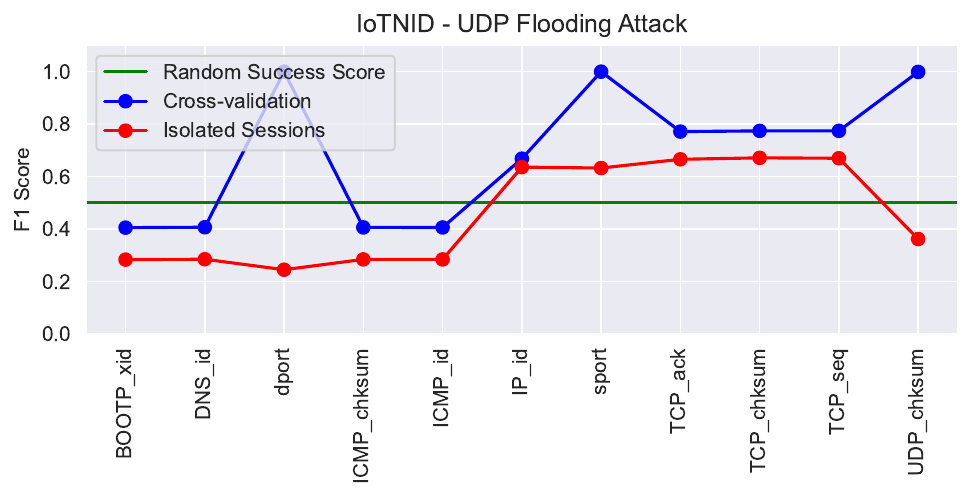}}}
		\label{fig:s1}	
\end{figure*}

In Section\ref{why} and \ref{lr} we argued that IPF-based attack detection yielded unrealistic results based on information leakage. To substantiate our claim, we conducted experiments using the IoT-NID~\cite{IoTNID} dataset, which offers multiple sessions for each attack. This dataset records each attack execution as a separate session, providing an opportunity to explore the identifiability of session-based features. We utilised examples from this dataset to demonstrate the effectiveness of session-based identifiers. Our approach involved merging the first and second sessions of the attacks to create a new dataset. We then performed a 10-times 10-fold CV on this merged dataset, employing one of the session-based identifiers at each iteration to distinguish between benign and attack samples. In the subsequent step, we executed the separation by utilising the first session as the training set and the second session as the testing set. Figure~\ref{fig:s1} illustrates the distribution of these features for four attacks.

The results of this analysis reveal a notable pattern: in the majority of cases, the identifying features demonstrate better success in the CV scenario, while in isolation, they exhibit limited effectiveness. Figure~\ref{fig:s1}  visually depicts this trend. In this context, these results support the hypothesis that these features lead to information leakage. 

\begin{figure}[htbp]
	\centering
	\caption{Visualization of high-achieving decision tree models. (a) HTTP Flood model using dport feature, (b) Brute-Force  model using dport feature. }
	\subfloat[\label{fig:HTTP_dport}\centering ]{{\includegraphics[width=60mm]{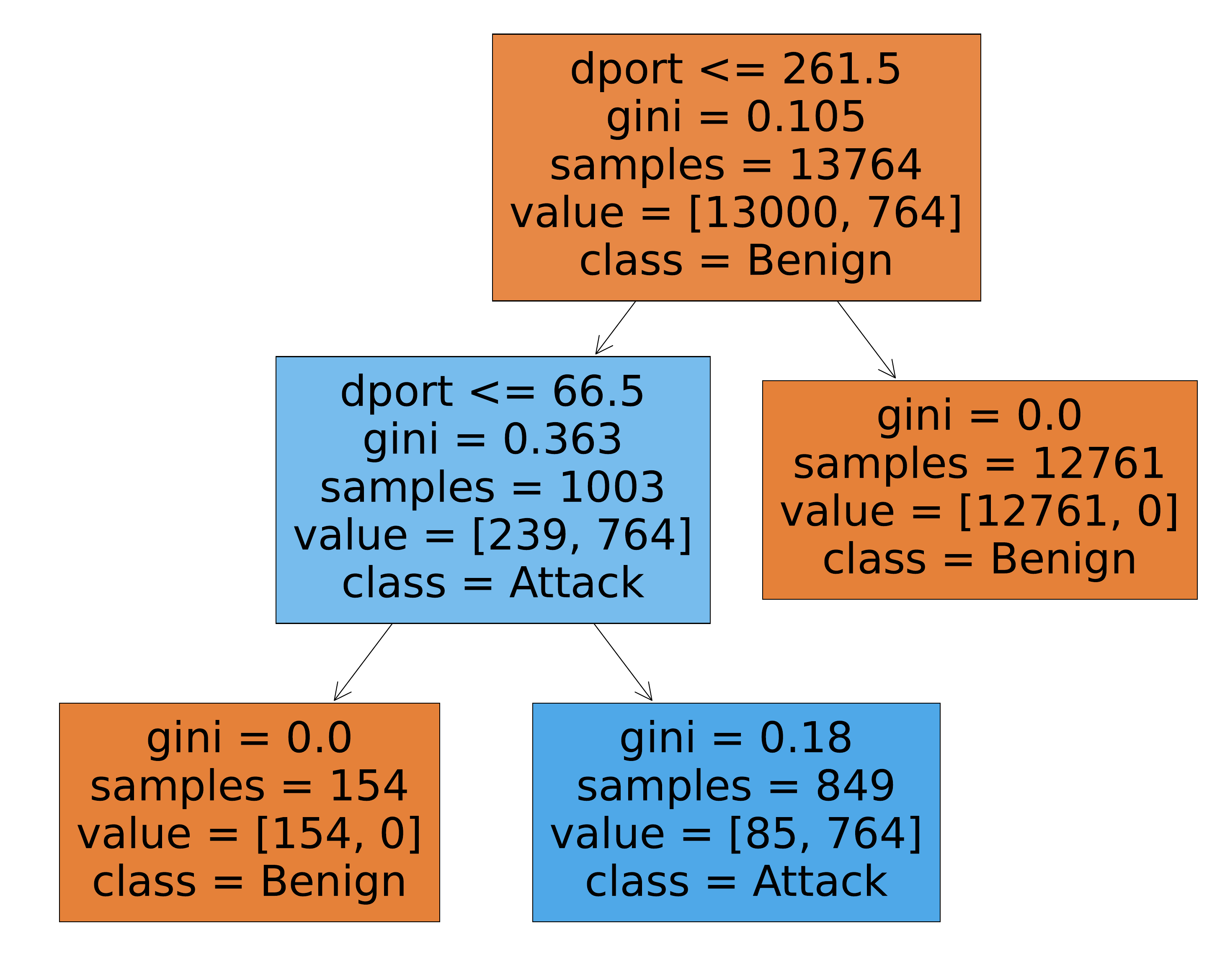}}}\\
	\subfloat[\label{fig:BF_dport}\centering ]{{\includegraphics[width=60mm]{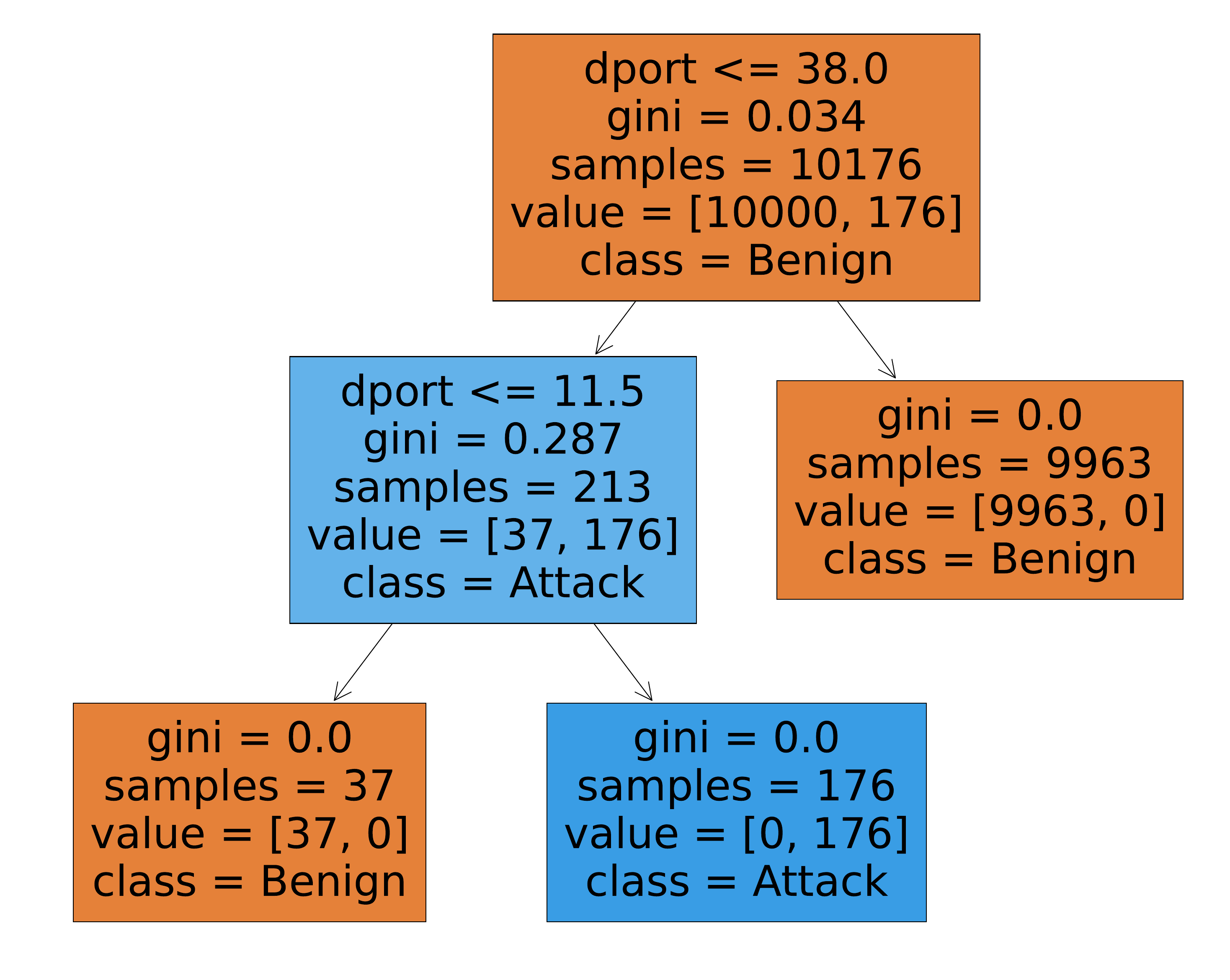}}}\\
	
	\label{fig:attackmodels}	
\end{figure}

However, upon a more comprehensive examination of the figures, we observed that certain features consistently display exceptional performance in both scenarios. Dport feature in Telnet Brute-force (Figure~\ref{fig:attack2}) and HTTP Flooding (Figure~\ref{fig:attack4})  attacks are prime examples of this consistent success across the two scenarios. If we examine the models associated with the features that achieve this unexpected success, as shown in Figure~\ref{fig:attackmodels}, we see that these models have an unusually simple structure. This brings us to the second explanation for achieving high performance with individual packet features: low data complexity.

\subsection{Low Data Complexity}\label{low}

\begin{figure}[htbp]
	\centering
	\caption[The distribution of packet size malicious data (HTTP Flood).]{The distribution of packet size malicious data (HTTP Flood) in four different datasets~\cite{koroniotis2019towards,CIC,ferrag2022edge,IoTNID}.}
	\centering{\includegraphics[width=80mm]{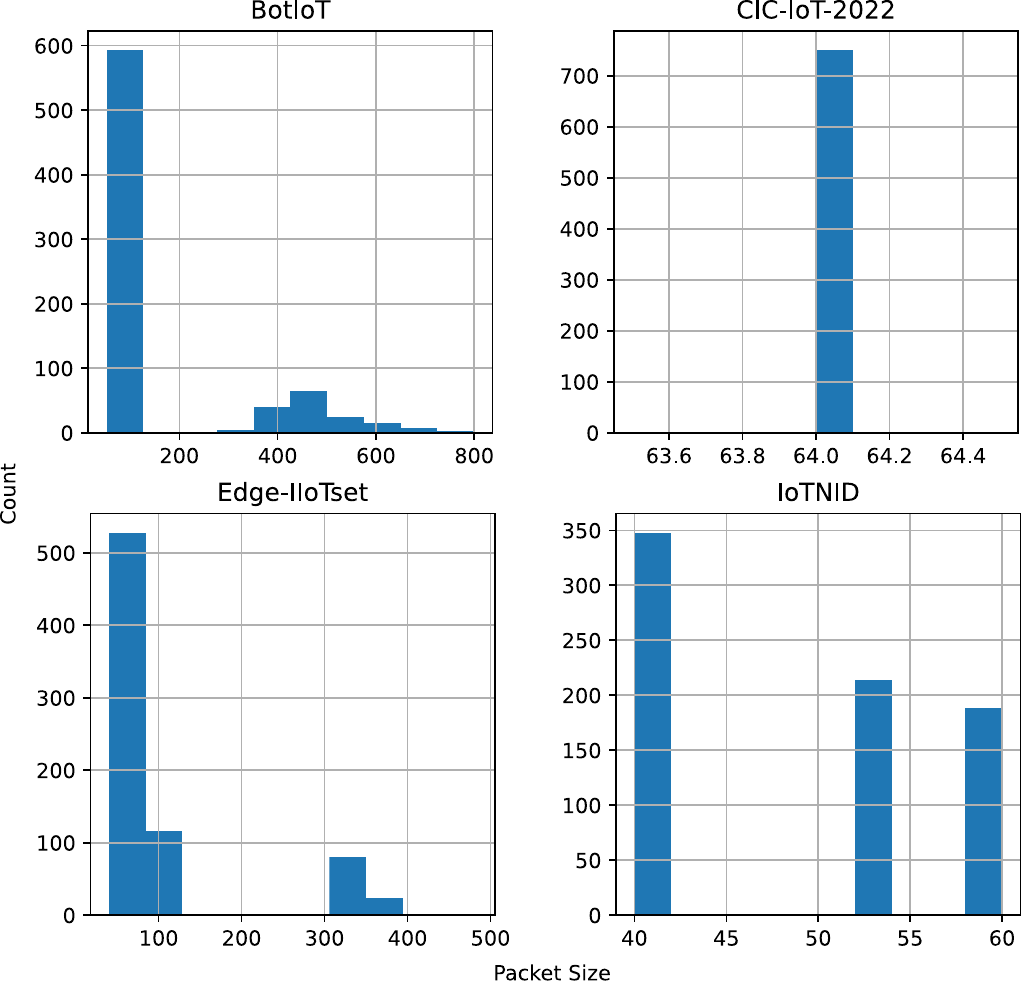}}%
	\label{fig:httphist}
\end{figure}

%Apart from some specific attacks such as MitM, 
Most attacks follow a pattern and produce uniform outputs. For example, the distribution of the size of the attack packets in the HTTP flood attack on four different datasets~\cite{koroniotis2019towards,CIC,ferrag2022edge,IoTNID} is shown in Figure~\ref{fig:httphist}. Because of the uniform nature of the attacks, if the complexity level of the dataset is low in studies using individual packets, even very basic characteristics, such as size or time, can become identifying features. 

\begin{figure}[htbp]
	\centering
	\caption{Distribution of labels in Kitsune and IoT-NID.
(a), (b), (c), and (d) show the size histograms of sessions 1, 2, 3, and 4 for UDP attack in the IoT-NID dataset, respectively.
(e) and (f) depict the distribution of labels over time for Mirai and Video Injection attacks, respectively, in the Kitsune.
 }
	\subfloat[\label{fig:comp_model1}\centering ]{{\includegraphics[width=40mm]{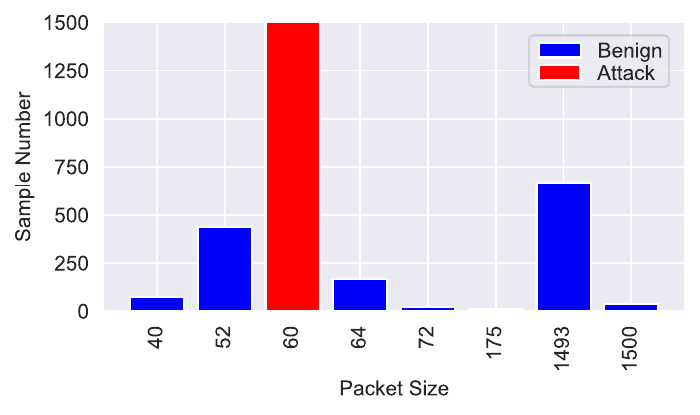}}}%
	\subfloat[\label{fig:comp_model2}\centering ]{{\includegraphics[width=40mm]{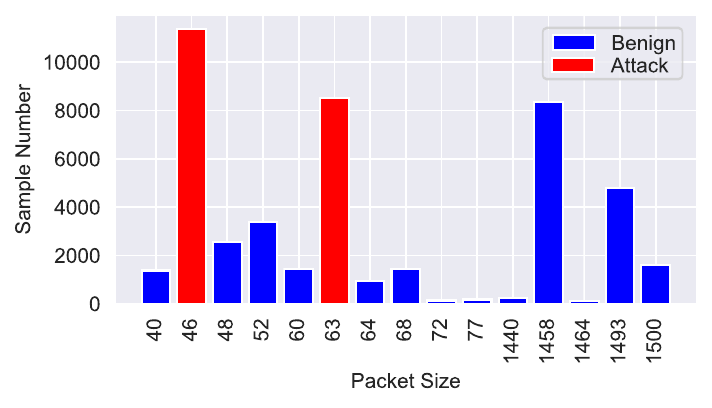}}}\\
	\subfloat[\label{fig:comp_model3}\centering ]{{\includegraphics[width=40mm]{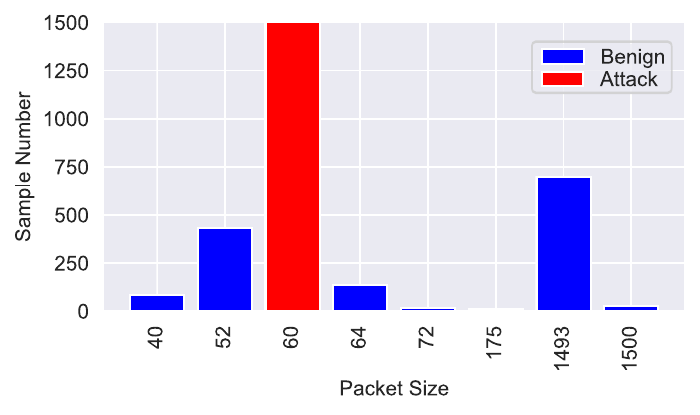}}}%
	\subfloat[\label{fig:comp_model4}\centering ]{{\includegraphics[width=40mm]{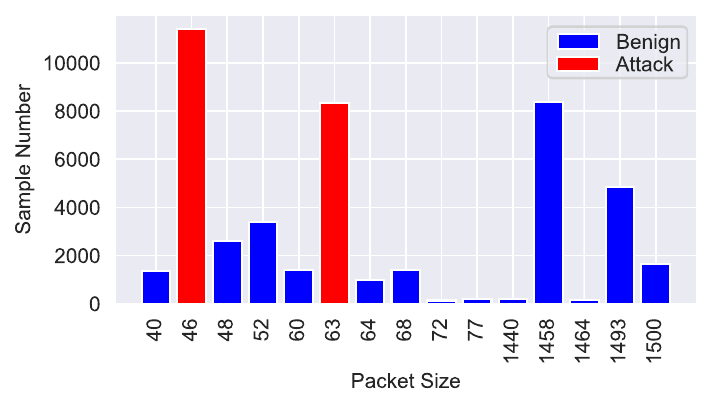}}}\\
	
	\subfloat[\label{fig:comp_model5}\centering]{{\includegraphics[width=80mm]{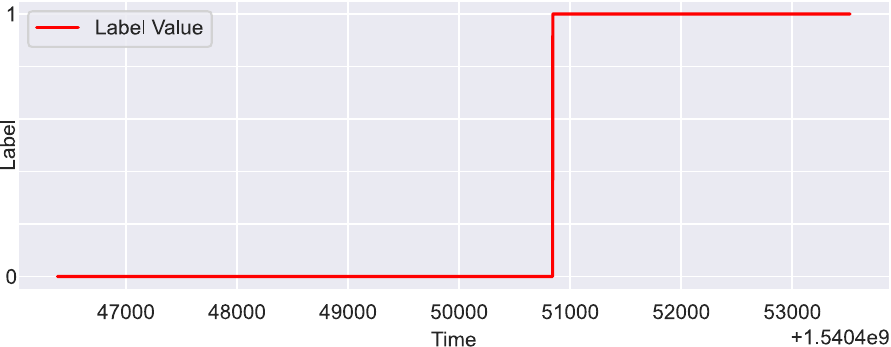}}}\\
	\subfloat[\label{fig:comp_model6}\centering]{{\includegraphics[width=80mm]{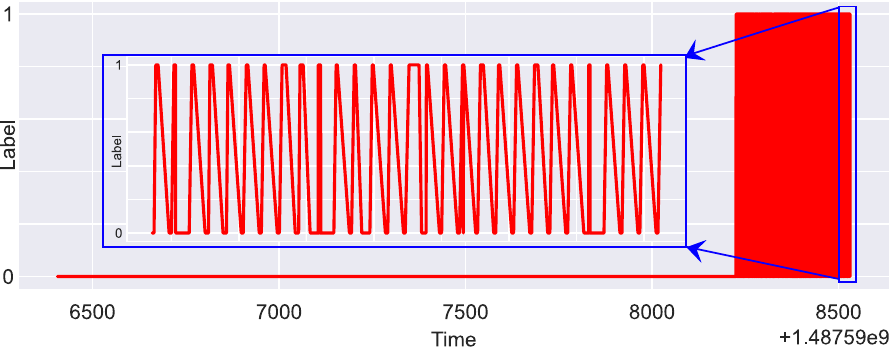}}}
	\label{fig:comp_model}	
\end{figure}

Figure~\ref{fig:comp_model} presents compelling empirical evidence to support this assertion. Specifically, in Figures~\ref{fig:comp_model1}-\subref{fig:comp_model4}, we depict the UDP attacks observed in four distinct sessions. A detailed analysis of these figures reveals notable similarities between sessions 1 (\ref{fig:comp_model1})  and 3 (\ref{fig:comp_model3}), as well as between sessions 2 (\ref{fig:comp_model2})  and 4 (\ref{fig:comp_model4}). Moreover, it is noteworthy that the size of attack packets remains constant even across these different sessions.

When incorporating only packet size as a feature in this example, we achieve perfect detection performance (see Table~\ref{tab:ind-results}). Even when testing the model trained with the first session on the third session, where the model has not encountered this data previously, we observe a notably high success rate. The same holds true for sessions 2 and 4, indicating strong consistency in their detection capabilities.  However, this superior success is only a consequence of the information leakage problem. The packet size in a UDP attack may change in another attack, and this model will fail to catch these attacks, or it will detect benign packets of this specific size as attacks.

Figure~\ref{fig:comp_model5} provides another insightful example, showcasing the change of labels over time in the Kitsune dataset~\cite{mirsky2018kitsune}, Mirai attack. In this graph, packets before a specific time exhibit all benign characteristics, while those occurring after the designated time are identified as attacks.

Additionally, we present yet another example, Figure~\ref{fig:comp_model6},  in which all packets up to a certain time demonstrate benign behaviour, while thereafter, a mixture of attack and benign packets is evident. The graph illustrates a dense concentration of data over a brief temporal window. However, focusing on a smaller portion of this mixture (the last 2000 samples) reveals a rhythmic pattern between attack and benign traffic over time.
The models trained using the time feature for these two attacks perform well on these datasets, but are not capable of detecting the same attack on another dataset or in real-life.

\begin{figure}
	\centering
	\caption[Complexity analysis of UDP Flood, Mirai, and Video injection attacks.]{Complexity analysis of UDP,  Mirai, and Video injection attacks according to time and size characteristics. (a) and (b) Mirai, (c) and (d) UDP, (e) and (f) Video injection time and size complexity, respectively. The central number depicted in the figures represents the comprehensive complexity score, which is derived as the average of 22 distinct analytical methods. (The complexity score ranges from 0 to 100, where higher values indicate increasing complexity). These 22 methods stem from six different approaches to assessing complexity, each distinctly colour-coded for clarity of categorisation~\cite{lorena2019complex}: red for feature-based, orange for linearity-based, yellow for neighbourhood-based, light green for network-based, dark green for dimensionality-based, and blue for class imbalance-based.}
	\subfloat[\label{fig:comp_t1}\centering ]{{\includegraphics[width=30mm]{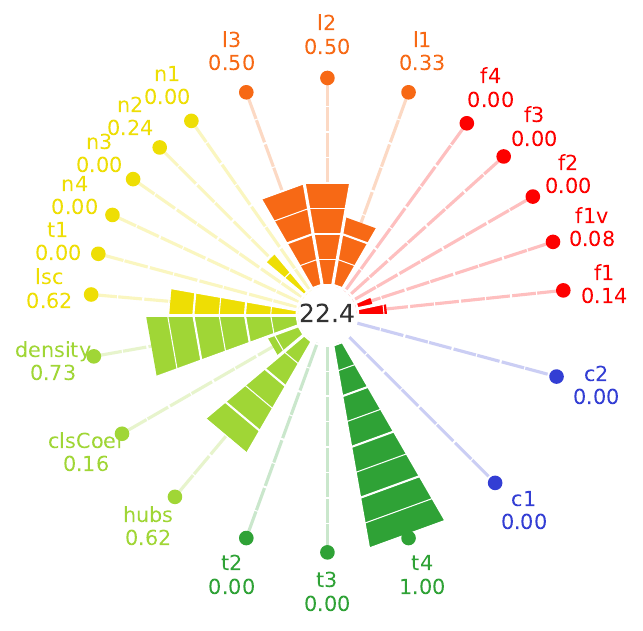}}}%
	\subfloat[\label{fig:comp_s1}\centering]{{\includegraphics[width=30mm]{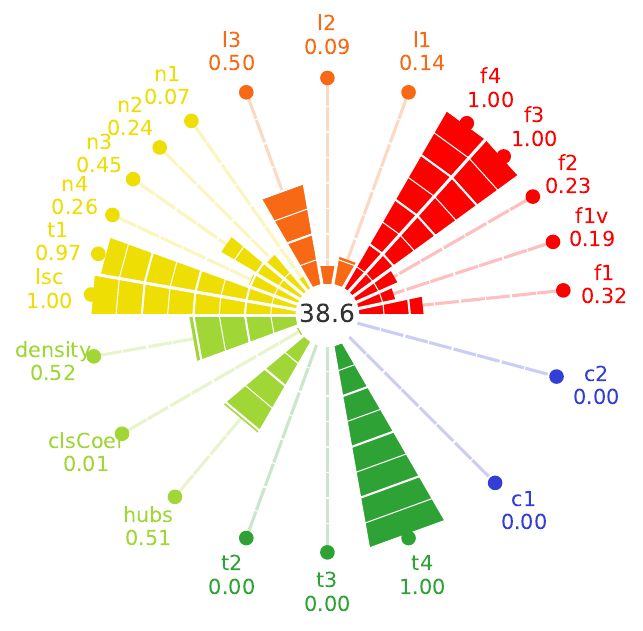}}}
	\subfloat[\label{fig:comp_t2}\centering]{{\includegraphics[width=30mm]{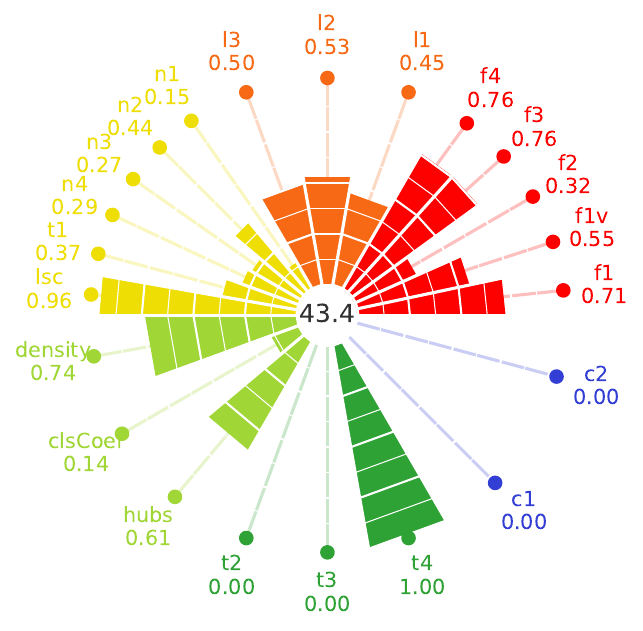}}}\\\subfloat[\label{fig:comp_s2}\centering]{{\includegraphics[width=30mm]{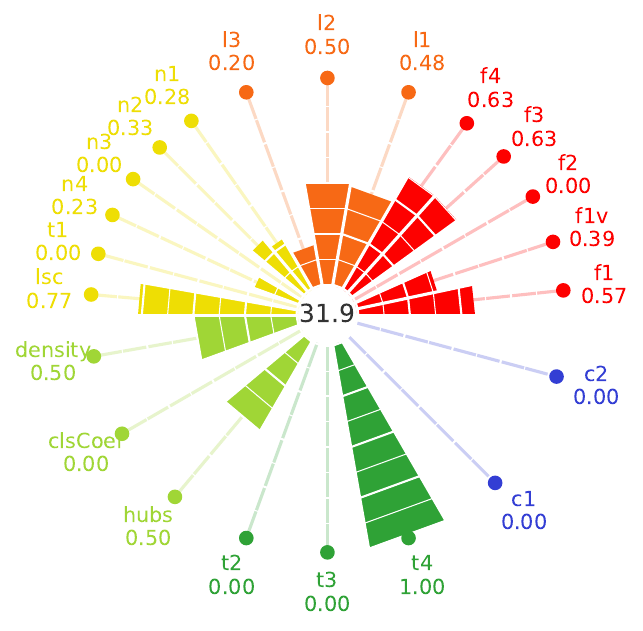}}}
	\subfloat[\label{fig:comp_t3}\centering]{{\includegraphics[width=30mm]{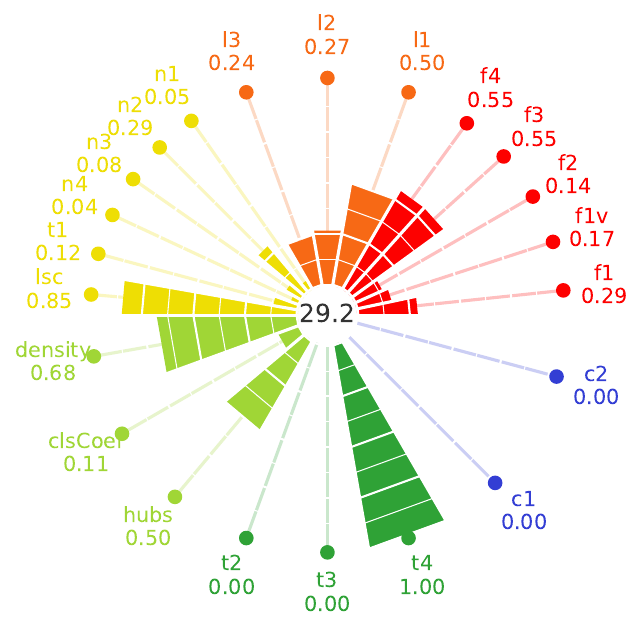}}}
	\subfloat[\label{fig:comp_s3}\centering]{{\includegraphics[width=30mm]{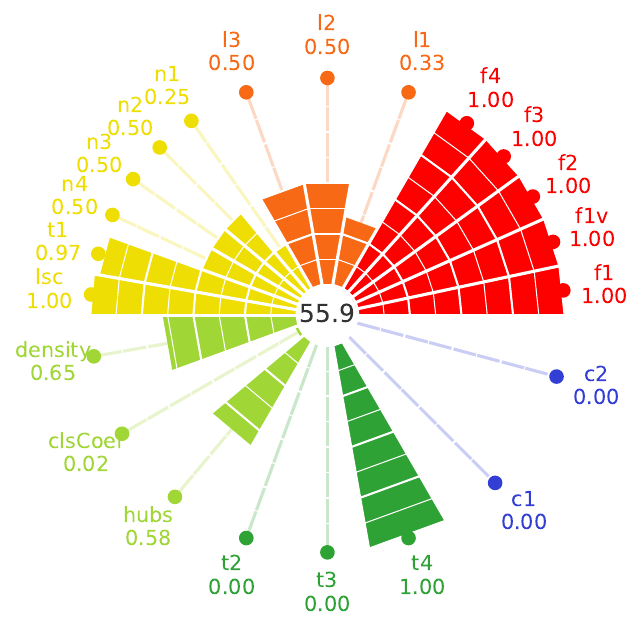}}}\\
	\label{fig:comp_graph}	
	
\end{figure}

To get a better understanding of the complexity of the data, we calculated the complexity score~\cite{lorena2019complex} of the attack files. Figure~\ref{fig:comp_graph} shows the computation of the complexity scores of 3 different data using time and size characteristics.
When these figures are analysed, it can be seen that the UDP flood attack has low size complexity and high time complexity. On the other hand, Mirai and video injection attacks show low size complexity and high time complexity.

\begin{table*}[htbp]
	\centering
	\caption[Near-perfect classification of 2 datasets afflicted by low data complexity.]{Near-perfect classification of 2 datasets afflicted by low data complexity. In Kitsune Mirai and Video injection data, the labels follow a certain time pattern. In the IoT-NID UDP data, attack packets have a specific size. In the Cross-Validation part of the table, the results depict the mean and standard deviation based on 10 repetitions of 10-fold cross-validation (using DT). In the isolated part, the mean and standard deviation are presented from 100 repetitions. Temporal information is not used in order to keep the features at the simplest level. The ``time'' feature here refers to a timestamp, not a temporal feature.}
	\resizebox{1.\textwidth}{!}{\begin{tabular}{@{}cllllrrrrr@{}}
			\toprule
			& Dataset & Subdataset & Feature & ML    & \multicolumn{1}{l}{Accuracy} & \multicolumn{1}{l}{Precsion} & \multicolumn{1}{l}{Recall} & \multicolumn{1}{l}{F1 Score} & \multicolumn{1}{l}{Kappa} \\
			\midrule
			\multirow{6}[2]{*}{\begin{sideways}Cross-validated\end{sideways}} & Kitsune & Video Inj & Size  & DT    & 0.959±0.000 & 0.479±0.000 & 0.500±0.000 & 0.489±0.000 & 0.000±0.000 \\
			& Kitsune & Mirai & Size  & DT    & 0.899±0.001 & 0.804±0.002 & 0.918±0.001 & 0.842±0.002 & 0.688±0.003 \\
			& IoT-NID & UDP-S2 & Size  & DT    & 1.000±0.000 & 1.000±0.000 & 1.000±0.000 & 1.000±0.000 & 0.999±0.001 \\
			& Kitsune & Video Inj & Time  & DT    & 0.988±0.000 & 0.933±0.002 & 0.919±0.002 & 0.926±0.001 & 0.852±0.003 \\
			& Kitsune & Mirai & Time  & DT    & 0.997±0.000 & 0.998±0.000 & 0.990±0.001 & 0.994±0.000 & 0.988±0.001  \\
			& IoT-NID & UDP-S2 & Time  & DT    & 0.841±0.007 & 0.833±0.008 & 0.839±0.008 & 0.835±0.008 & 0.671±0.016 \\
			\midrule
			\multirow{4}[2]{*}{\begin{sideways}Isolated\end{sideways}} & IoT-NID & UDP-S2vsS3 & Size  & DT    & 0.032±0.000 & 0.016±0.000 & 0.500±0.000 & 0.031±0.000 & 0.000±0.000 \\
			& IoT-NID & UDP-S2vsS4 & Size  & DT    & 1.000±0.000 & 1.000±0.000 & 1.000±0.000 & 1.000±0.000 & 0.999±0.000 \\
			& IoT-NID & UDP-S2vsS3 & Time  & DT    & 0.526±0.000 & 0.508±0.000 & 0.566±0.000 & 0.379±0.000 & 0.017±0.000 \\
			& IoT-NID & UDP-S2vsS4 & Time  & DT    & 0.866±0.000 & 0.858±0.000 & 0.865±0.000 & 0.861±0.000 & 0.722±0.000 \\
			\bottomrule
	\end{tabular}}%
	\label{tab:ind-data}%
\end{table*}%

Table~\ref{tab:ind-data} shows the results of the attack detection process using only one feature for each of the three attacks with shared complexity values. Upon examination of the CV (10-times, 10-fold) results in Table~\ref{tab:ind-data}, we observe that the UDP attack, characterised by very low complexity in terms of size, achieves flawless detection. Similarly, in the Mirai attack, very low time complexity leads to nearly perfect detection rates. Conversely, the video injection attack, which possesses a relatively higher size complexity but lower time complexity, did not achieve substantial success in terms of size-based detection but demonstrated significant success when considering time-based detection.

In our second experiment, capitalising on the multi-session nature of the UDP attack, We conducted tests involving the second session, pairing it with the fourth session, which has very similar characteristics. Additionally, we tested the second session with the third sessions, which are notably different from it. The similarities and differences between these sessions can be better understood by examining their distributions in Fig~\ref{fig:comp_model}. The results, shown under ``isolated" in Table~\ref{tab:ind-data}, indicate exceptional performance in the fourth sessions that were similar, while not achieving notable results in the third sessions exhibited substantial dissimilarities from the second session.

What all of these examples have in common is that high levels of success can be achieved by using individual features that are not suitable for attack detection. Although identifying features are not used here, the uniformity of attack data enables even basic individual characteristics to serve as identifying features.

\begin{figure}[htbp]
	\centering	\caption[Comparison of the change in the complexity value and model success.]{Reflection of the change in the complexity value due to a single feature on the model success (with DT).}
	\includegraphics[width=80mm]{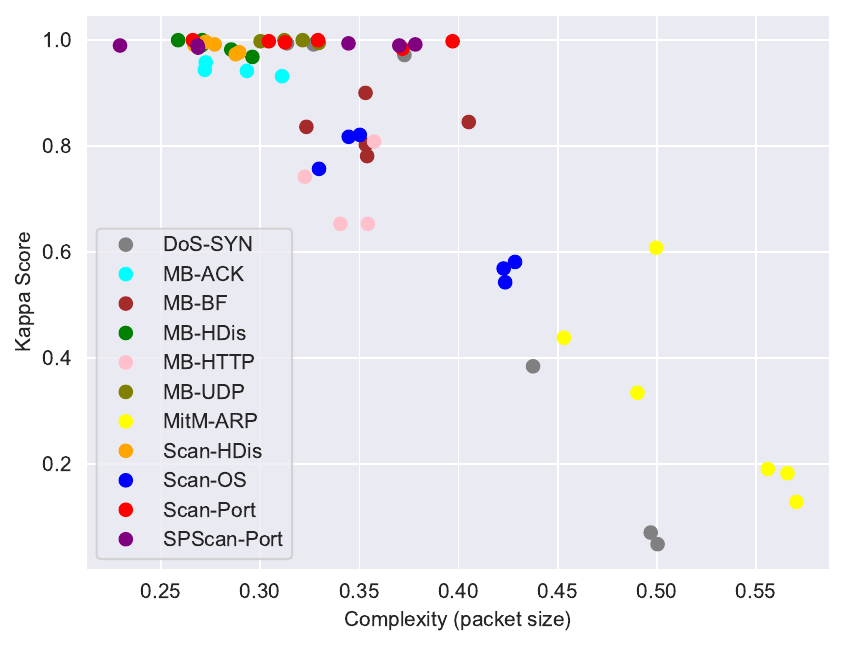}
	
	\label{fig:compexity_graph}
\end{figure}

To give more insight into the relationship between complexity and overfitting, we have expanded this sample a little more. Figure~\ref{fig:compexity_graph} shows the results of another experiment on complexity. In this experiment, a basic feature (packet size) was used to separate all attack types in the IoT-NID dataset. From this graph, we can observe the inverse correlation between complexity and success. Especially, if the complexity is below the critical level (45), highly successful discrimination is achieved. On the other hand, as complexity increases, success seems to decrease.

\section{Conclusion} 

In summary, this study highlights the limitations of relying solely on individual packet features (IPF) for intrusion detection within IoT environments. While conducted within the IoT domain to emphasise security considerations in IoT devices, the implications extend beyond this realm. Our findings stress the necessity of holistic approaches considering contextual features and packet interactions for effective intrusion detection in various network security contexts.

Through literature review and experimental analysis, we've shown that although IPF may show high detection rates, they suffer from inherent flaws like information leakage and low data complexity. This underscores the importance of prioritising robust ML-based intrusion detection systems that address the multifaceted nature of security challenges in networked environments.

Moving forward, researchers and practitioners should prioritise comprehensive approaches incorporating contextual features and packet interactions to enhance IoT ecosystem resilience and networked environment security against evolving cyber-threats. By addressing these limitations and embracing sophisticated detection methodologies, we can strengthen defences and mitigate risks posed by malicious actors.

\bibliographystyle{IEEEtran}
\bibliography{references}

% that's all folks
\end{document}